\documentstyle[12pt]{article}

\begin{document}

\def\reff#1{(\ref{#1})}
\newcommand{\csd}{critical slowing-down}
\newcommand{\be}{\begin{equation}}
\newcommand{\ee}{\end{equation}}
\newcommand{\<}{\langle}
\renewcommand{\>}{\rangle}
\newcommand{\half}{ {{1 \over 2 }}}
\newcommand{\quarter}{ {{1 \over 4 }}}
\newcommand{\fourth}{\quarter}
\newcommand{\eighth}{ {{1 \over 8 }}}
\newcommand{\sixteenth}{ {{1 \over 16 }}}
\def\var{ \hbox{var} }
\newcommand{\HB}{ {\hbox{{\scriptsize\em HB}\/}} }
\newcommand{\MGMC}{ {\hbox{{\scriptsize\em MGMC}\/}} }
\newcommand{\gtilde}{ {\widetilde{G}} }
\newcommand{\longto}{\longrightarrow}

\def\spose#1{\hbox to 0pt{#1\hss}}
\def\ltapprox{\mathrel{\spose{\lower 3pt\hbox{$\mathchar"218$}}
 \raise 2.0pt\hbox{$\mathchar"13C$}}}
\def\gtapprox{\mathrel{\spose{\lower 3pt\hbox{$\mathchar"218$}}
 \raise 2.0pt\hbox{$\mathchar"13E$}}}

\newcommand{\scra}{{\cal A}}
\newcommand{\scrb}{{\cal B}}
\newcommand{\scrc}{{\cal C}}
\newcommand{\scrd}{{\cal D}}
\newcommand{\scre}{{\cal E}}
\newcommand{\scrf}{{\cal F}}
\newcommand{\scrg}{{\cal G}}
\newcommand{\scrh}{{\cal H}}
\newcommand{\scrk}{{\cal K}}
\newcommand{\scrl}{{\cal L}}
\newcommand{\scrm}{{\cal M}}
\newcommand{\scrmvec}{\vec{\cal M}}
\newcommand{\scrp}{{\cal P}}
\newcommand{\scrr}{{\cal R}}
\newcommand{\scrs}{{\cal S}}
\newcommand{\scru}{{\cal U}}

\def\bsigma{\mbox{\protect\boldmath $\sigma$}}

\title{Comment on \\ ``Antiferromagnetic Potts Models''}

\author{
  \small  Mona Lubin      \\[-0.2cm]
  \small\it Center for Neural Science     \\[-0.2cm]
  \small\it New York University       \\[-0.2cm]
  \small\it 2--4 Washington Place        \\[-0.2cm]
  \small\it New York, NY 10003 USA    \\[-0.2cm]
  \small Internet:  {\tt MONA@CNS.NYU.EDU} \\[-0.2cm]
    \\ \and
  \small  Alan D. Sokal               \\[-0.2cm]
  \small\it Department of Physics     \\[-0.2cm]
  \small\it New York University       \\[-0.2cm]
  \small\it 4 Washington Place        \\[-0.2cm]
  \small\it New York, NY 10003 USA    \\[-0.2cm]
  \small Internet:  {\tt SOKAL@ACF3.NYU.EDU} \\[-0.2cm]
  {\protect\makebox[5in]{\quad}}  
   \\
}

\vspace{0.2cm}
\maketitle
\thispagestyle{empty}   

\vspace{-0.4cm}
\begin{abstract}
We show that the Wang-Swendsen-Koteck\'y algorithm
for antiferromagnetic $q$-state Potts models
is nonergodic at zero temperature for $q=3$
on periodic $3m \times 3n$ lattices
where $m,n$ are relatively prime.
For $q \ge 4$ and/or other lattice sizes or boundary conditions,
the ergodicity at zero temperature is an open question.
\end{abstract}

\clearpage

%
%

Wang, Swendsen and Koteck\'y (WSK) \cite{WSK}
have recently proposed an elegant Monte Carlo algorithm
for simulating the antiferromagnetic $q$-state Potts model
on a finite graph $G$.
It goes as follows:
Choose at random two distinct ``colors'' $\alpha,\beta \in \{1,\ldots,q\}$;
freeze all the spins taking values $\neq \alpha,\beta$,
and allow the remaining spins to take value either $\alpha$ or $\beta$.
The induced model is then an antiferromagnetic Ising model,
which can be updated by any legitimate algorithm
(for example, the Swendsen-Wang algorithm \cite{Swendsen-Wang}
 or Wolff's single-cluster variant \cite{Wolff_89a}).

At zero temperature the antiferromagnetic $q$-state Potts model
reduces to the equal-weight distribution on $q$-colorings of $G$,
and the WSK algorithm becomes:  independently for each connected cluster of
$\alpha-\beta$ spins, either leave that cluster as is
or else flip it (interchanging $\alpha$ and $\beta$).

WSK studied their algorithm numerically for
(among other cases) the $q=3$ model at $T=0$
on square lattices of linear size $L=4,8,16,32,64$
with periodic boundary conditions.   
They claimed that the autocorrelation time was $\tau_{WSK} \approx 7$
independent of $L$,
while the autocorrelation time of a single-spin-flip algorithm
increased approximately as $\tau_{SSF} \approx 0.32 L^2$.

If the (exponential) autocorrelation time of a Monte Carlo algorithm
is {\em finite}\/, then in particular that algorithm must be ergodic.
However, WSK did not give any proof of the ergodicity of their
algorithm at $T=0$.  (The ergodicity at $T \neq 0$ is trivial.)
Here we show that in fact the algorithm is {\em not}\/ ergodic at $T=0$
for $q=3$ on periodic lattices of size $3m \times 3n$
where $m,n$ are relatively prime.
For $q \ge 4$ and/or other lattice sizes or boundary conditions,
the ergodicity at $T=0$ is an open question.

\begin{figure}[b]
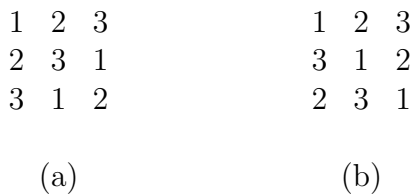

\begin{center}
$$
\begin{array}{ccc}
   \begin{array}{ccc}
      1 & 2 & 3 \\
      2 & 3 & 1 \\
      3 & 1 & 2
   \end{array}
   &  \qquad\qquad   &
   \begin{array}{ccc}
      1 & 2 & 3 \\
      3 & 1 & 2 \\
      2 & 3 & 1
   \end{array}
   \\
                &  &              \\
   \hbox{(a)}   &  & \hbox{(b)}
\end{array}
$$
\end{center}
\caption{
   Configurations of the zero-temperature antiferromagnetic 3-state Potts
   model on a $3 \times 3$ lattice with periodic boundary conditions.
}
\label{fig1}
\end{figure}

Consider the configurations shown in Figure \ref{fig1}
for a $3 \times 3$ periodic lattice.
For any choice of $\alpha,\beta$,
the sites colored $\alpha-\beta$ form a {\em single}\/ connected cluster,
so the only possible moves in the WSK algorithm are global permutations
of the colors.
On the other hand, configurations (a) and (b) are {\em not}\/
related by a global permutation,
since in (a) the bands of constant color run northeast-southwest
while in (b) they run northwest-southeast.
It follows that the WSK algorithm is nonergodic.

Next consider the configurations of Figure \ref{fig1}
repeated periodically on a $3m \times 3n$ lattice.
If $m,n$ are relatively prime,
then the sites colored $\alpha-\beta$ form a {\em single}\/ connected band
winding around the lattice,
and the argument goes through unchanged.
If $m,n$ are not relatively prime,
then the sites colored $\alpha-\beta$ form {\em several}\/ disjoint
connected bands,
and the ergodicity is an open problem.
%
%
%
%

We remark that these configurations are completely frozen
under any single-spin-update algorithm,
because each spin is surrounded by neighbors of both colors.
So any such algorithm is also nonergodic.
The same holds for $q=4$ on lattices $4m \times 4n$,
and for $q=5$ on lattices $5m \times 5n$.
For $q \ge 6$,
the single-spin-update algorithm is easily seen to be ergodic
on any square lattice \cite{Jerrum_private};
more generally, this holds on an arbitrary graph $G$
for $q \ge \max\deg G + 2$.

We also remark that the WSK algorithm for $q=3$ is nonergodic
on the {\em planar}\/ graph of \cite[Fig.~10.4]{Even_book}.

To see these nonergodicities {\em numerically}\/ requires some care:

(a)  One must study the model not only {\em at}\/ $T=0$,
but also at temperatures $T$ {\em approaching}\/ zero;
then one will see the autocorrelation time growing without limit.

(b)  One must measure an observable that distinguishes between the
ergodic classes.  For example, in the above situation one could use
$|\widetilde{\bsigma}(k)|^2$ at momenta $k = (2\pi/3, \pm 2\pi/3)$.
It is not clear which observables might be sensitive to any possible
nonergodicities at other values of $q$ and $L$.

It is an open question whether there exist efficient algorithms
for simulating the antiferromagnetic Potts model at zero temperature
for $q=3,4,5$ (more generally, for $q < \max\deg G + 2$).
Jerrum \cite{Jerrum_private} has pointed out that it is unlikely
that such algorithms (with polynomially bounded autocorrelation time
measured in CPU units) can exist for arbitrary graphs $G$
and {\em fixed}\/ $q$:
indeed, the existence of such an algorithm for $q=5$
would permit one to ascertain with high probability
the 3-colorability of an arbitrary degree-4 graph,
which is impossible if NP $\neq$ RP \cite{NP_RP}.

We wish to thank Mark Jerrum and Greg Sorkin for helpful correspondence.

\clearpage


\begin{thebibliography}{99}

\bibitem{WSK}  J.-S. Wang, R.H. Swendsen and R. Koteck\'y,
  Phys. Rev. Lett. {\bf 63}, 109 (1989)
  and Phys. Rev. {\bf B42}, 2465 (1990).

\bibitem{Swendsen-Wang}  R.H. Swendsen and J.-S. Wang,
  Phys. Rev. Lett. {\bf 58}, 86 (1987).

\bibitem{Wolff_89a}  U. Wolff, Phys. Rev. Lett. {\bf 62}, 361 (1989).

\bibitem{Jerrum_private}  M. Jerrum, private communication.

\bibitem{Even_book} S. Even, {\em Graph Algorithms}\/
   (Computer Science Press, Potomac MD, 1979), p.~220.

\bibitem{NP_RP}  For the definitions of NP and RP, see
M.R. Garey and D.S. Johnson, {\em Computers and Intractability}\/
(Freeman, San Francisco, 1979)
and D.J.A. Welsh, Discrete Appl. Math. {\bf 5}, 133 (1983).


%
%
%

\end{thebibliography}
\end{document}